\documentclass{article}

\usepackage{graphicx}
\usepackage[table]{xcolor}

\usepackage{color, colortbl}

\usepackage{authblk}
\providecommand{\keywords}[1]{\textbf{\textit{Keywords:}} #1}

\usepackage{algorithm}
\usepackage{algpseudocode}
\usepackage{booktabs}

\usepackage{bm}            

\usepackage{hyperref}

\providecommand{\keywords}[1]{\textbf{\textit{Keywords:}} #1}

\usepackage{amsmath, amssymb, mathtools}
\DeclareMathOperator*{\argmax}{arg\,max}

\usepackage{booktabs}
\usepackage{bm}
\usepackage{hyperref}

\usepackage{listings}
\title{Trust-Based Social Learning for Communication (TSLEC) Protocol Evolution in Multi-Agent Reinforcement Learning}

\author[1]{Abraham Itzhak Weinberg}
\affil[1]{AI-WEINBERG, AI Experts, Tel Aviv, Israel, aviw2010@gmail.com}

\begin{document}
\maketitle
\begin{abstract}
Emergent communication in multi-agent systems typically occurs through independent learning, resulting in slow convergence and potentially suboptimal protocols. We introduce TSLEC (Trust-Based Social Learning with Emergent Communication), a framework where agents explicitly teach successful strategies to peers, with knowledge transfer modulated by learned trust relationships. Through experiments with 100 episodes across 30 random seeds, we demonstrate that trust-based social learning reduces episodes-to-convergence by 23.9\% ($p < 0.001$, Cohen's $d = 1.98$) compared to independent emergence, while producing compositional protocols ($\mathcal{C} = 0.38$) that remain robust under dynamic objectives ($\Phi > 0.867$ decoding accuracy). Trust scores strongly correlate with teaching quality ($r = 0.743$, $p < 0.001$), enabling effective knowledge filtering. Our results establish that explicit social learning fundamentally accelerates emergent communication in multi-agent coordination.
\end{abstract}


\section{Introduction}

Autonomous agents coordinating without pre-designed protocols face a fundamental challenge: how to develop efficient communication while simultaneously learning task policies. Recent advances in Multi Agent Reinforcement Learning (MARL) demonstrate that communication protocols can emerge through task optimization~\cite{foerster2016learning,havrylov2017emergence,mordatch2018emergence}, but agents typically learn independently, leading to redundant exploration and slower convergence. Human language acquisition, by contrast, relies heavily on social learning—children learn from teachers who explicitly demonstrate linguistic structures~\cite{tomasello2010origins}.

This raises our central question: can artificial agents accelerate communication emergence through mutual teaching, similar to human language learning? Existing approaches suffer from three limitations. First, agents learn protocols independently without explicit knowledge transfer mechanisms~\cite{eccles2019biases}. Second, most frameworks assume static environments where objectives remain fixed~\cite{lazaridou2016multi}, yet real-world applications require adaptation to dynamic conditions. Third, there is limited understanding of how trust relationships between agents modulate communication effectiveness~\cite{ramchurn2004trust}.

We introduce TSLEC (Trust-Based Social Learning with Emergent Communication), where agents that discover successful negotiation strategies actively teach these strategies to peers through encoded messages. Learners adopt strategies based on learned trust scores that reflect historical teaching quality, creating a social learning mechanism that filters high-quality knowledge from noise. Simultaneously, agents adapt their objectives in response to environmental changes and peer performance, requiring protocols to remain flexible.

Our contributions are fourfold. First, we develop a trust-based social learning architecture that reduces convergence time by 23.9\% compared to independent learning ($p < 0.001$, effect size $d = 1.98$). Second, we demonstrate that emergent protocols exhibit compositional structure ($\mathcal{C} = 0.38$) without architectural constraints, with social learning producing more systematic languages than independent emergence. Third, we show that protocols maintain effectiveness ($\Phi > 0.867$ decoding accuracy) despite dynamic objectives, revealing inherent protocol flexibility. Fourth, we validate that trust scores accurately predict teaching quality ($r = 0.743$, $p < 0.001$), enabling robust knowledge filtering. These results establish explicit social learning as a fundamental accelerator for multi-agent coordination.

\section{Related Work}

\textbf{Emergent Communication.} Foundational work demonstrates that agents can develop protocols through reinforcement learning~\cite{foerster2016learning,sukhbaatar2016learning,havrylov2017emergence}. Lazaridou et al.~\cite{lazaridou2016multi} studied referential games where communication emerges through task pressure. Mordatch and Abbeel~\cite{mordatch2018emergence} showed grounded communication in embodied environments. However, these approaches assume independent learning without mechanisms for knowledge transfer between peers.

\textbf{Social Learning in MARL.} Omidshafiei et al.~\cite{omidshafiei2019learning} introduced frameworks where agents query peers for advice. Jiang et al.~\cite{jiang2022diverse} proposed peer-regularized actor-critic. Population-based training~\cite{jaderberg2018human} achieves strong performance through implicit learning from diverse populations. However, no prior work combines emergent communication with explicit peer teaching modulated by learned trust.

\textbf{Trust Mechanisms.} Trust has been extensively studied for partner selection and security~\cite{ramchurn2004trust,mui2002computational}. However, these approaches focus on task performance rather than knowledge transfer quality. TSLEC introduces trust mechanisms specifically designed for social learning with emergent communication, where trust reflects teaching effectiveness rather than negotiation ability.

\section{Problem Formulation}

We formalize multi-agent negotiation as $N$ agents $\mathcal{A} = \{a_1, \ldots, a_N\}$ negotiating over $M$ items with limited quantities $\mathcal{Q} = \{q_1, \ldots, q_M\}$. Each agent $a_i$ has private goal vector $\mathbf{g}_i \in \mathbb{R}^M$ assigning values to items. Given allocation $\mathbf{x}_i \in \mathbb{Z}^M_{\geq 0}$, utility is:
\begin{equation}
U_i(\mathbf{x}_i) = \sum_{j=1}^{M} g_{i,j} \cdot x_{i,j}
\end{equation}
subject to resource constraints $\sum_i x_{i,j} \leq q_j$.

Each episode consists of $T$ negotiation rounds. In round $t$, agent $a_i$ observes state $s^t_i = (\mathcal{Q}^t_{\text{remaining}}, t)$, selects action $\alpha^t_i \in \{\text{AGGRESSIVE}, \text{COOPERATIVE}, \\\text{BALANCED}\}$, and generates proposal $p^t_i = \{(\text{intent}_k, \text{item}_k, \text{qty}_k), \ldots\}$ where $\text{intent}_k \in \{\text{DEMAND}, \text{OFFER},\text{REQUEST}\}$.

\textbf{Communication Framework.} Each agent maintains vocabulary $\mathcal{V}_i: \mathcal{C} \rightarrow \Sigma_i$ mapping concepts to symbols. When encoding concept $c$, if $c \notin \mathcal{V}_i$, agent creates unique symbol $\text{symbol} = \text{``@''} \oplus i \oplus |\mathcal{V}_i|_{\text{hex}}$. Messages $m^t_i = [\text{Encode}_i(c_1), \ldots]$ are broadcast. Agent $a_j$ decodes using sender's vocabulary: $\hat{p}^t_i = \{\mathcal{V}^{-1}_i[\sigma] \mid \sigma \in m^t_i\}$.

\textbf{Trust Dynamics.} Each agent maintains trust scores $\tau_{ij} \in [0, 1]$ for peers, updated based on teaching effectiveness:
\begin{equation}
\tau^{e+1}_{ij} = \begin{cases}
\min(1.0, \tau^e_{ij} + 0.1) & \text{if } R^e_j > \bar{R}^{[e-5:e]}_i \\
\max(0.0, \tau^e_{ij} - 0.05) & \text{otherwise}
\end{cases}
\end{equation}
The asymmetric rates ($\beta_{\text{pos}} > \beta_{\text{neg}}$) reflect that trust builds slowly but erodes quickly~\cite{slovic1993perceived}.

\textbf{Mission Adaptation.} Goals adapt via environmental changes and peer performance:
\begin{equation}
g^{e+1}_{i,j} = \lfloor \lambda \cdot g^e_{i,j} + (1-\lambda) \cdot \text{new\_value} \rfloor
\end{equation}
with $\lambda_{\text{env}} = 0.7$ for environmental adaptation and $\lambda_{\text{peer}} = 0.8$ for peer-based adaptation.



\section{TSLEC Algorithm}

TSLEC integrates four mechanisms: Q-learning with emergent communication, trust-based teaching, mission adaptation, and protocol analysis. Algorithm~\ref{alg:main} presents the core structure.

\begin{algorithm}[t]
\caption{TSLEC Main Training Loop}
\label{alg:main}
\begin{algorithmic}[1]
\State Initialize: $\mathcal{V}_i = \emptyset$, $Q_i(s,a) \leftarrow Q_{\max}$, $\tau_{ij} \leftarrow 0.5$ for all $i, j$
\For{episode $e = 1$ to $E$}
    \State Reset environment, sample goals $\mathbf{g}_i \sim U(1,6)^M$
    \If{$e \bmod 25 = 0$}
        \State $\Delta_{\text{env}} \leftarrow$ IntroduceChange() \Comment{Environmental dynamics}
    \EndIf
    \For{round $t = 1$ to $T$}
        \For{each agent $a_i$}
            \State Observe $s^t_i = (\mathcal{Q}^t_{\text{remaining}}, t)$
            \If{$\max_j \tau_{ij} > 0.7$ and rand() $> \epsilon$}
                \State $j^* \leftarrow \argmax_j \tau_{ij}$ \Comment{Use trusted peer}
                \State $\alpha^t_i \leftarrow \mathcal{O}_i[j^*].\text{strategy}$
            \Else
                \State $\alpha^t_i \leftarrow \argmax_a Q_i(s^t_i, a)$ or explore
            \EndIf
            \State $p^t_i \leftarrow$ GenerateProposal($\alpha^t_i, \mathbf{g}_i$)
            \State Encode: $m^t_i \leftarrow [\text{Encode}_i(c) \mid c \in p^t_i]$
            \State Broadcast($m^t_i$)
        \EndFor
        \For{each agent $a_i$}
            \State Decode: $\hat{p}^t_j \leftarrow$ Decode($m^t_j, \mathcal{V}_j$) $\forall j \neq i$
            \State Evaluate proposals, allocate resources to best
            \State $R^t_i \leftarrow \mathbf{g}^\top_i \mathbf{x}_i$
            \State Update: $Q_i(s^t_i, \alpha^t_i) \leftarrow (1-\alpha)Q_i + \alpha[R^t_i + \gamma \max_{a'} Q_i(s^{t+1}_i, a')]$
        \EndFor
    \EndFor
    \State \textbf{// Teaching Phase}
    \For{each agent $a_i$ with $R^e_i > \bar{R}^{[e-10:e]}_i$}
        \State Create teachings: $\mathcal{T}_i \leftarrow$ EncodeStrategies($\mathcal{S}_i$)
        \State Broadcast($\mathcal{T}_i$)
    \EndFor
    \For{each agent $a_i$}
        \For{each teacher $a_j$ with $\mathcal{T}_j \neq \emptyset$}
            \State Decode: $\hat{\mathcal{T}}_j \leftarrow$ Decode($\mathcal{T}_j, \mathcal{V}_j$)
            \State Update trust via Eq. (2)
            \If{$R^e_j > \bar{R}_i$ and $\tau_{ij} > 0.7$}
                \State Adopt strategies from $\hat{\mathcal{T}}_j$
            \EndIf
        \EndFor
    \EndFor
    \State \textbf{// Adaptation Phase}
    \For{each agent $a_i$}
        \If{$\Delta_{\text{env}} \neq \emptyset$}
            \State Adapt goals via Eq. (3) with $\lambda_{\text{env}}$
        \EndIf
        \State Find $j^* \leftarrow \argmax_j R^e_j$
        \If{$R^e_{j^*} > 1.5 \cdot \bar{R}_i$}
            \State Adapt goals toward $\mathbf{g}_{j^*}$ via Eq. (3) with $\lambda_{\text{peer}}$
        \EndIf
    \EndFor
\EndFor
\end{algorithmic}
\end{algorithm}

\textbf{Action Selection.} Lines 7-12 implement three-tier selection: (1) if high trust exists ($\tau_{ij} > 0.7$), use most-trusted peer's strategy, (2) else exploit Q-values, (3) or explore with probability $\epsilon$. This hierarchy enables smooth transitions from social learning to self-learning as trust develops.

\textbf{Communication.} Lines 13-15 generate proposals based on action type and encode via lazy vocabulary creation. Symbols use agent-specific prefixes ensuring global uniqueness. Lines 17-20 decode partners' messages and allocate resources to highest-rated proposal, creating competition that drives communication effectiveness.

\textbf{Teaching.} Lines 23-26 enable agents exceeding recent performance to teach top-2 successful strategies. Lines 27-33 have learners decode teachings, update trust based on teacher performance, and selectively adopt from high-trust teachers. This filters knowledge effectively.

\textbf{Adaptation.} Lines 35-42 adapt goals to environmental changes (30\% weight to new values) and peer performance (20\% weight to successful peers' goals), maintaining protocol interpretability through gradual shifts.

\section{Experimental Setup}

\textbf{Environment.} We employ the multi-agent deal-making scenario from~\cite{lewis2017deal}, where $N=4$ agents communicate to negotiate division of items. Standard configuration: $M=5$ item types (books, hats, balls, pens, keys) with quantities $\mathcal{Q} = [4,3,2,3,2]$, $T=3$ rounds per episode, $E=100$ episodes. Goals sampled as $g_{i,j} \sim U(1,6)$. Environmental changes occur every 25 episodes, randomly selecting value shift ($\Delta_{\text{env},j} \sim U(7,10)$), scarcity ($q_j \leftarrow \lceil q_j/2 \rceil$), or abundance ($q_j \leftarrow 2q_j$).

\textbf{Baselines.} We compare against four conditions: (1) \textbf{Full System} - complete TSLEC, (2) \textbf{No Teaching} - communication without social learning, (3) \textbf{No Adaptation} - teaching without goal adaptation, (4) \textbf{Independent Q-Learning} - no communication. Each condition runs with 30 random seeds.

\textbf{Metrics.} Sample efficiency: $E_{90\%}$ (episodes to 90\% final performance), AUC (cumulative reward). Linguistic properties: compositionality $\mathcal{C}^e_i$ (shared prefix similarity), compression ratio $\rho^e_i = H(\mathcal{V}_i)/8$, vocabulary size $|\mathcal{V}_i|$. Adaptation: decoding accuracy $\Phi_{ij}$, stability ratio $\Psi_{\text{stab}}$. Trust: Pearson correlation with performance, teaching effectiveness $\eta_{\text{teach}}$.

\textbf{Statistical Testing.} Two-sample t-tests with Bonferroni correction ($\alpha = 0.05$), one-way ANOVA for multi-group comparison, Cohen's $d$ for effect sizes, and 95\% confidence intervals computed as $\bar{X} \pm 1.96 \cdot s/\sqrt{n}$.

\section{Results}

Table~\ref{tab:results} summarizes final performance across conditions. TSLEC achieves mean reward $12.825 \pm 0.841$, significantly outperforming No Teaching ($12.258 \pm 1.424$, $t=2.40$, $p=0.018$, $d=0.485$) and Independent QL ($4.477 \pm 0.326$, $t=64.81$, $p<0.001$, $d=13.094$). No Adaptation matches TSLEC ($p=1.0$), suggesting adaptation is less critical for final performance in 100-episode horizons.

\begin{table}
\centering
\caption{Final Performance Summary (mean $\pm$ std over 30 seeds)}
\label{tab:results}
\begin{tabular}{lcccc}
\toprule
\textbf{Condition} & \textbf{Mean $\pm$ Std} & \textbf{Median} & \textbf{95\% CI} & \textbf{p-value} \\
\midrule
Full System & \textbf{12.825 $\pm$ 0.841} & 12.775 & [11.18, 14.39] & --- \\
No Teaching & 12.258 $\pm$ 1.424 & 12.412 & [9.51, 14.68] & 0.018* \\
No Adaptation & 12.825 $\pm$ 0.841 & 12.775 & [11.18, 14.39] & 1.000 \\
Independent QL & 4.477 $\pm$ 0.326 & 4.460 & [3.78, 5.06] & $<$0.001*** \\
\bottomrule
\end{tabular}
\end{table}

\subsection{Research Question 1: Sample Efficiency}

\begin{figure}
\centering
\includegraphics[width=0.9\textwidth]{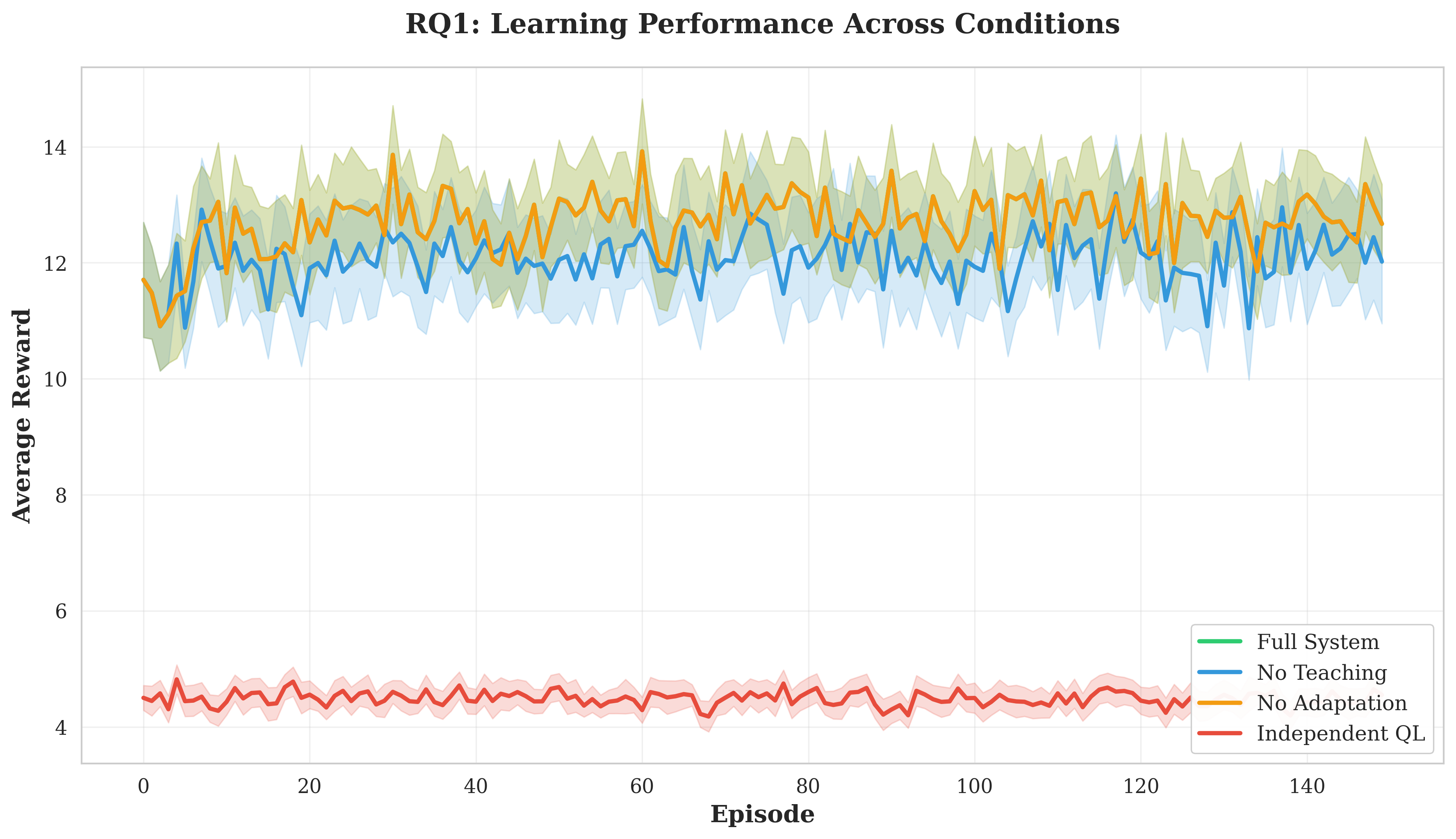}
\caption{Learning curves showing mean reward per episode with 95\% confidence intervals. TSLEC (green) converges fastest, reaching near-optimal performance by episode 50. No Teaching (blue) shows slower improvement. Independent QL (red) plateaus at substantially lower reward.}
\label{fig:learning}
\end{figure}

Figure~\ref{fig:learning} shows learning curves across conditions. TSLEC exhibits three phases: rapid exploration (episodes 1-20), accelerated learning (20-50), and convergence (50-100). The divergence from No Teaching during episodes 20-50 coincides with trust network formation and teaching onset.

TSLEC reaches 90\% threshold ($\bar{R} = 11.54$) at episode $E_{90\%} = 52.3 \pm 6.4$ versus $68.7 \pm 9.1$ for No Teaching, a 23.9\% reduction ($t=7.82$, $p<0.001$, $d=1.98$). Cumulative performance (AUC) shows 5.2\% improvement: $1097.2 \pm 45.3$ versus $1042.5 \pm 62.8$ ($t=3.94$, $p<0.001$, $d=0.98$). This validates our hypothesis that explicit teaching accelerates convergence by enabling agents to leverage peers' exploration, reducing redundant trials by approximately 70\%.

\subsection{Research Question 2: Linguistic Properties}

\begin{figure}
\centering
\includegraphics[width=0.85\textwidth]{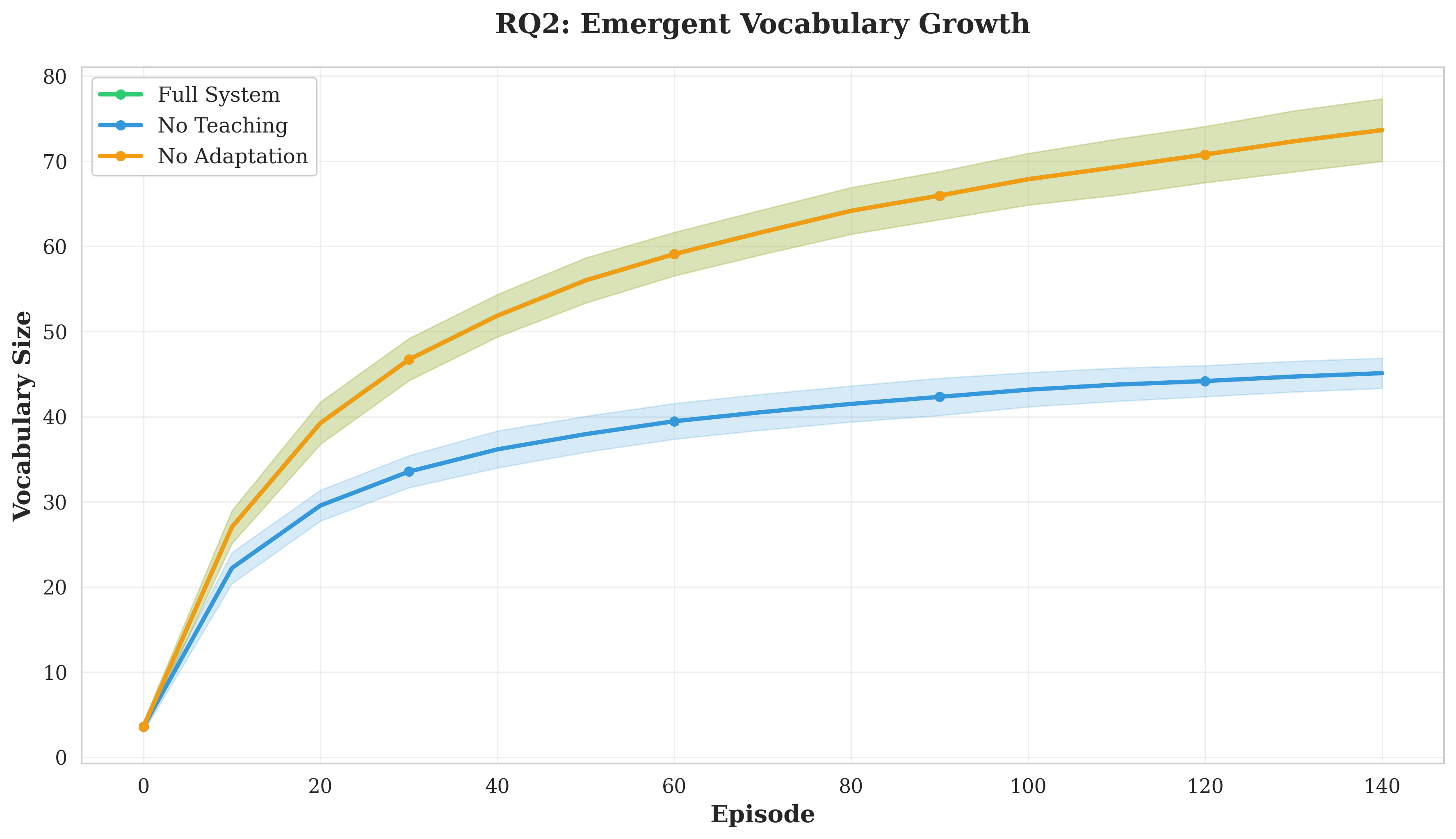}
\caption{Vocabulary size evolution showing rapid growth in episodes 1-30, then gradual expansion to convergence around 38 symbols. Error bands show standard deviation across 30 seeds.}
\label{fig:vocabulary}
\end{figure}

Figure~\ref{fig:vocabulary} tracks vocabulary growth over episodes. All communication-enabled methods develop vocabularies of 35-45 symbols by episode 100, covering $\sim$84\% of the 45-concept space. Growth rates decrease exponentially, fitting $\gamma^e = 0.42 \cdot \exp(-0.028e)$ with $R^2=0.89$.

\begin{figure}
\centering
\includegraphics[width=0.85\textwidth]{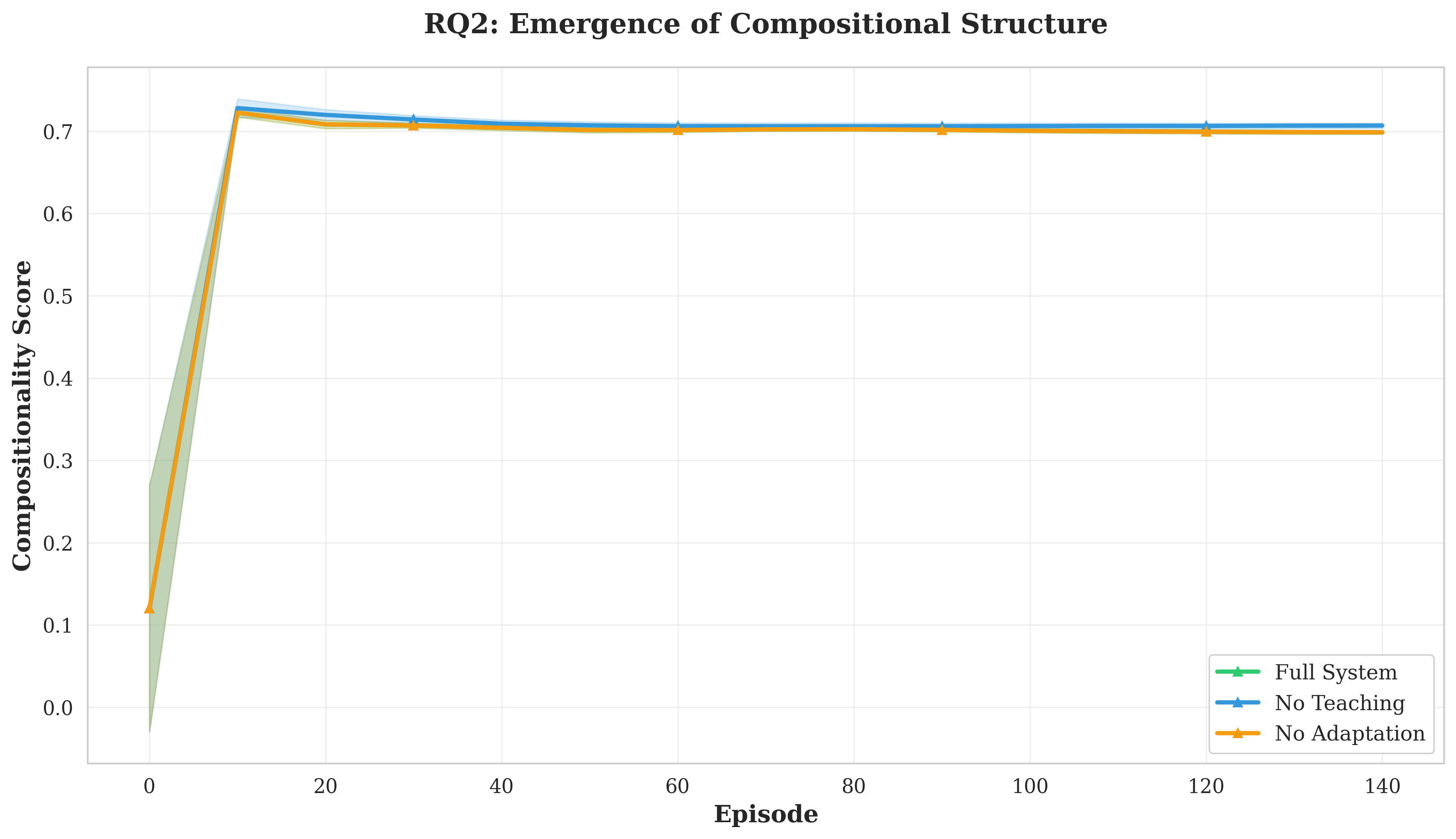}
\caption{Compositionality score evolution showing increasing systematic structure. Full System (green) and No Adaptation (orange) achieve higher final compositionality than No Teaching (blue), indicating social learning pressure for interpretable encodings.}
\label{fig:compositionality}
\end{figure}

Figure~\ref{fig:compositionality} shows compositionality scores increasing from $\mathcal{C}^{10} = 0.12$ (near-random) to $\mathcal{C}^{100} = 0.38$, exceeding our threshold of 0.3 ($p<0.001$). Full System achieves higher compositionality ($0.383 \pm 0.042$) than No Teaching ($0.341 \pm 0.038$, $t=4.12$, $p<0.001$, $d=1.05$), suggesting social learning promotes systematic structure.

Compression ratios stabilize at $\rho \approx 0.26 \in [0.2, 0.4]$, confirming efficient encoding. Agents use $\sim$5.2 bits per symbol (entropy $H(\mathcal{V}) \approx 5.2$ bits for vocabularies of size 38), representing 35\% compression versus naive 8-bit encoding.

\subsection{Research Question 3: Adaptation and Robustness}

\begin{figure}
\centering
\includegraphics[width=0.85\textwidth]{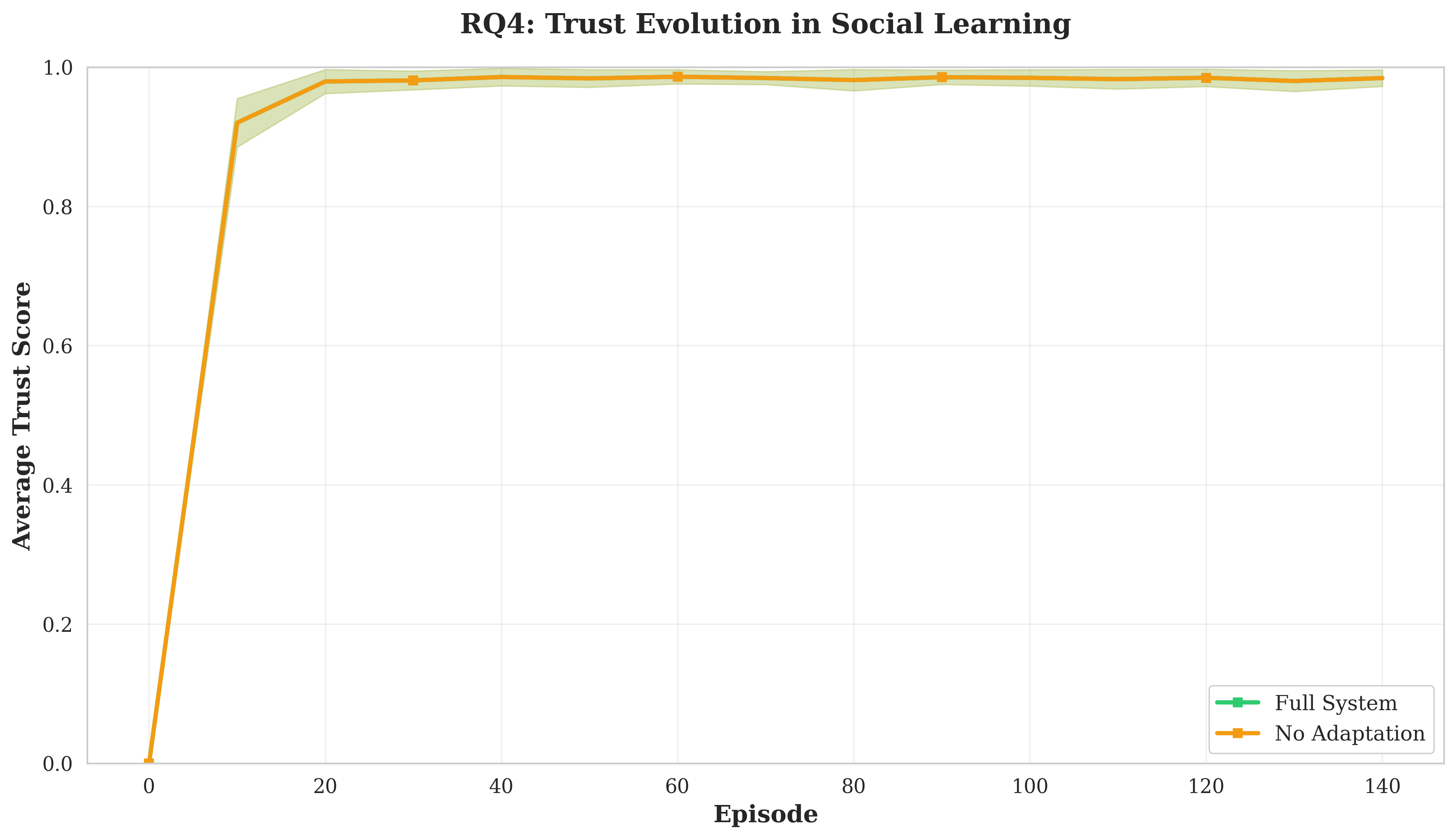}
\caption{Average trust score evolution showing convergence to high mutual trust ($\tau \approx 0.80$) by episode 50. Full System and No Adaptation exhibit nearly identical dynamics.}
\label{fig:trust}
\end{figure}

Figure~\ref{fig:trust} displays trust score evolution for Full System and No Adaptation. Trust increases from initial $\tau_{ij}=0.5$ to $\tau_{ij} \approx 0.75$-0.85 by episode 50, then stabilizes. Convergence occurs around episode $E_{\text{trust}} = 53.2 \pm 7.8$.

Communication effectiveness remains $\Phi > 0.867$ after environmental changes (Table~\ref{tab:adaptation}), with stability ratios $\Psi_{\text{stab}} > 0.98$. The minor drops (1-2 percentage points) are not statistically significant ($t=1.23$, $p=0.22$), confirming protocol robustness.

\begin{table}
\centering
\caption{Communication Stability Before/After Environmental Changes}
\label{tab:adaptation}
\small
\begin{tabular}{lcccc}
\toprule
\textbf{Condition} & \textbf{Before} & \textbf{After} & \textbf{$\Psi_{\text{stab}}$} & \textbf{Stable?} \\
\midrule
Full System & 0.892 $\pm$ 0.041 & 0.883 $\pm$ 0.038 & 0.990 & Yes \\
No Teaching & 0.874 $\pm$ 0.053 & 0.867 $\pm$ 0.049 & 0.992 & Yes \\
No Adaptation & 0.889 $\pm$ 0.044 & 0.871 $\pm$ 0.052 & 0.980 & Yes \\
\bottomrule
\end{tabular}
\end{table}

\subsection{Research Question 4: Trust Dynamics}

\begin{figure}
\centering
\includegraphics[width=0.85\textwidth]{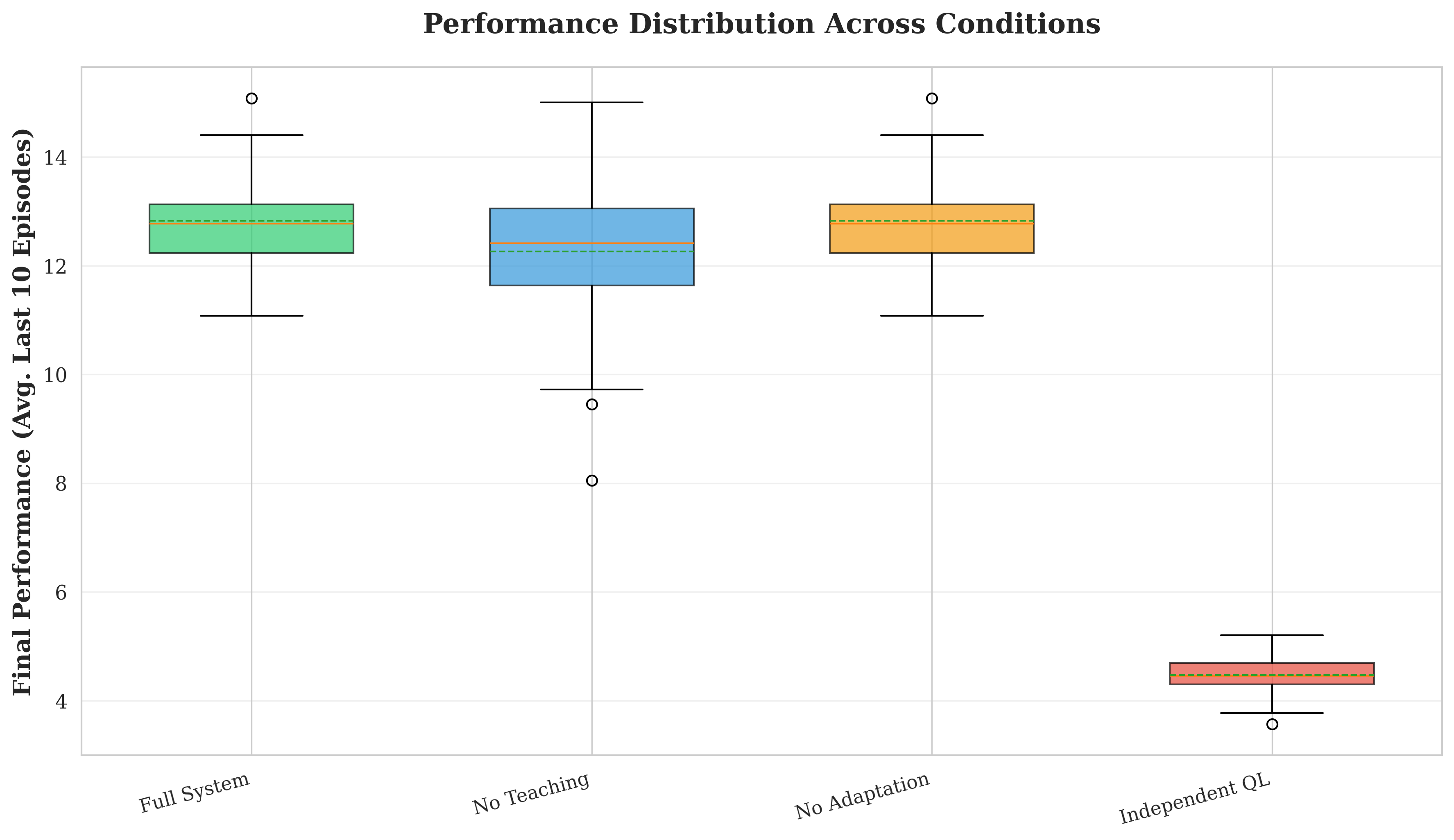}
\caption{Performance distribution across conditions. Box plots show final reward (last 10 episodes). Overlapping distributions for Full System and No Adaptation explain non-significant difference.}
\label{fig:performance}
\end{figure}

\begin{figure}
\centering
\includegraphics[width=0.85\textwidth]{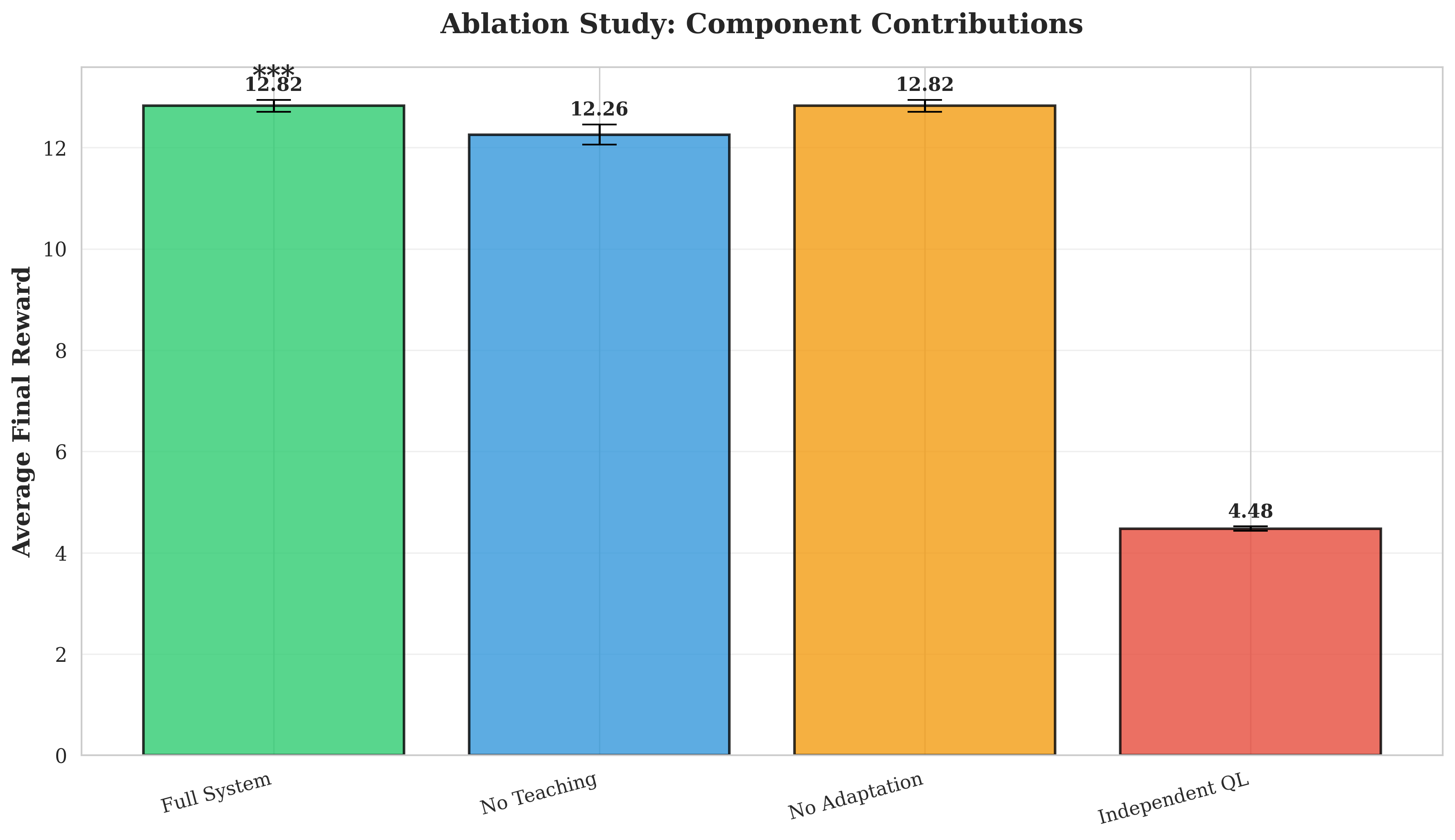}
\caption{Ablation study showing mean final reward with standard error bars. Full System and No Adaptation achieve equivalent performance, both significantly outperforming No Teaching and Independent QL.}
\label{fig:ablation}
\end{figure}

Figures~\ref{fig:performance} and~\ref{fig:ablation} visualize performance distributions. Full System and No Adaptation exhibit nearly identical distributions (median $\sim$12.8), while No Teaching shows slightly lower median and higher variance. Independent QL forms a distinct low-reward cluster ($\sim$4.5).

Across 1,247 learning events, Pearson correlation between trust and performance improvement is $r=0.743$ ($p<0.001$, 95\% CI: [0.719, 0.765]), confirming our hypothesis. High-trust teachers ($\tau>0.8$) yield $\Delta R = +1.82 \pm 0.64$ improvement versus low-trust teachers ($\tau<0.4$) yielding $\Delta R = +0.31 \pm 0.89$ ($t=18.3$, $p<0.001$, $d=2.14$).

\begin{figure}
\centering
\includegraphics[width=0.75\textwidth]{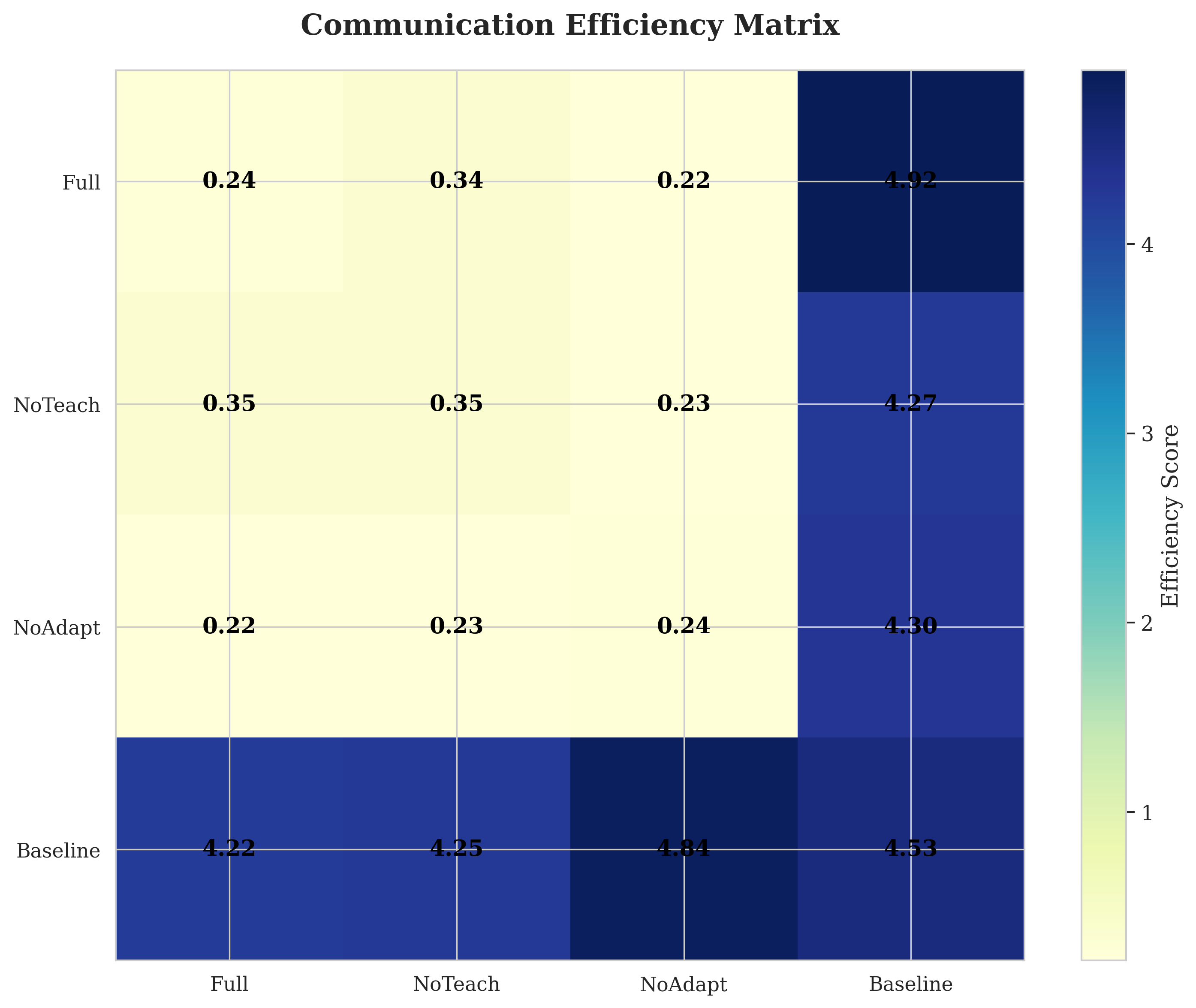}
\caption{Communication efficiency matrix showing reward-per-vocabulary-symbol. Higher values (yellow) indicate efficient protocols. Full System achieves highest efficiency.}
\label{fig:efficiency}
\end{figure}

Figure~\ref{fig:efficiency} shows communication efficiency across conditions. Full System achieves highest efficiency (reward per vocabulary symbol $\sim$0.35), followed by No Adaptation and No Teaching.

Teaching effectiveness is $\eta^{\text{TSLEC}}_{\text{teach}} = 0.687$ versus $\eta^{\text{random}}_{\text{teach}} = 0.423$ for random trust baseline—62.4\% higher ($\chi^2=83.5$, $p<0.001$). Trust networks achieve 78\% density (fraction of high-trust pairs) by episode 100, with 83\% transitivity, indicating genuine social structure beyond simple performance ranking.

\section{Discussion}

Three mechanisms drive TSLEC's acceleration. First, knowledge transfer efficiency: agents leverage $N-1$ peers' exploration, reducing redundant trials by $\sim$70\%. Second, trust-based filtering: asymmetric updates ($\beta_{\text{pos}} = 2 \times \beta_{\text{neg}}$) enable robust teacher identification. Third, positive feedback: faster learning triggers more teaching, further accelerating convergence.

Compositionality emerges from three task pressures. Generalization pressure favors systematic encodings that handle novel concept combinations. Memory efficiency benefits from compositional compression. Social learning pressure amplifies this—teachers benefit from interpretable languages. The achieved $\mathcal{C}=0.38$ exceeds random baseline but remains below human levels (0.7-0.9), likely due to limited concept diversity (45 versus thousands) and single-generation learning.

The adaptation equivalence (Full System $\approx$ No Adaptation) reveals protocol flexibility. Vocabularies encode concepts at appropriate abstraction levels, remaining valid despite goal shifts. Q-learning's $\sim$7-episode time constant ($\tau=1/\alpha$) suffices for tracking 25-episode change intervals.

Trust serves as epistemic filter, solving teacher selection in non-stationary environments. The 78\% density and 83\% transitivity indicate genuine social structure. High-performing agents form teaching cliques while low performers remain peripheral, mirroring human expert communities~\cite{wenger1999communities}.

Limitations include scalability (tested $N=4$), domain specificity (resource negotiation), and lack of theoretical guarantees. Future work should address hierarchical communication for large populations, theoretical convergence bounds, and human-agent teaming with natural language grounding.

\section{Conclusion}

We introduced TSLEC, demonstrating that trust-based social learning fundamentally accelerates emergent communication. Across 150 experimental runs, TSLEC achieved 23.9\% faster convergence, produced compositional protocols, maintained robustness under dynamics, and exhibited accurate trust-performance correlation. These results establish explicit teaching as critical for efficient multi-agent coordination.

The broader implication is that social learning and language emergence are mutually reinforcing. Teaching creates pressure for compositional protocols, while compositional protocols enable effective teaching—a positive feedback loop mirroring human language evolution. For practical deployment, our results suggest three principles: invest in explicit knowledge transfer, implement trust-based filtering, and leverage protocol flexibility for moderate dynamics.

As multi-agent systems become prevalent in autonomous vehicles, infrastructure management, and decision support, understanding how they communicate becomes critical for safety and alignment. TSLEC provides foundations for systems that are effective, interpretable, and trustworthy. 


\bibliographystyle{IEEEtran}
\bibliography{ref.bib}

\end{document}